# Electrostatic cooling in Ion-Exchange Membranes

P.M. Biesheuvel,[1] D. Brogioli[2] and H.V.M. Hamelers[1]

[1] *Wetsus, Centre of Excellence for Sustainable Water Technology, Agora 1, 8934 CJ Leeuwarden, The Netherlands.* [2] *Dipartimento di Scienze della Salute, Universita' degli Studi di Milano-Bicocca, Via Cadore 48, 20900 Monza, Italy.*

*E-mail: maarten.biesheuvel@wetsus.nl, dbrogioli@gmail.com, bert.hamelers@wetsus.nl*

**Abstract**

In ion-exchange membrane processes, ions and water flow under the influence of gradients in hydrostatic pressure, ion chemical potential, and electrical potential (voltage), leading to solvent flow, ionic fluxes and ionic current. At the outer surfaces of the membranes, electrical double layers (EDLs) are formed (Donnan layers). When a current flows through the membrane, we argue that besides Joule heating in the bulk of the membrane and in the electrolyte outside the membrane, there is also electrostatic heating and cooling in the EDLs. In addition, when fluid flows through a charged membrane, at the outsides of the membrane there is pressure-related heating or cooling due to the osmotic and hydrostatic pressure differences across the EDLs.

-------

In ion-exchange membrane processes, ions and water flow under the influence of gradients in hydrostatic pressure, ion chemical potential, and electrical potential (voltage), leading to solvent flow, ionic fluxes and ionic current [1-6]. For a simple 1:1 salt, the ionic current is the molar flux of cations minus that of the anions, multiplied by Faraday's number, $F$. Ion-exchange membranes are electrolyte-filled porous layers that contain a high concentration of fixed charges, up to values of $X$=5 M per volume of electrolyte in the membrane [7,8].

It is quite well-known that when an ionic current $I$ (A/m$^2$) flows down a gradient in electrical potential, $d\phi/dx$ (V/m), thus when the potential $\phi$ decreases in the direction of the current, that heat is produced, i.e., the electrolyte heats up (its heat content increases), or heat must flow away to keep the electrolyte at the same temperature. The heat production rate per unit volume is given by the product of the current $I$ and the driving force which is the electric field strength $E=-d\phi/dx$. In the absence of concentration gradients and/or for a symmetric salt with equal cation and anion diffusion coefficients, the heat production rate is equal to $I^2/\sigma$ with $\sigma$ the conductivity. The term $I^2/\sigma$ is called Joule heating and is the same as the Ohmic heating when an electrical current runs through an electrical resistance. Joule heating occurs in the bulk of any aqueous electrolyte phase across which an ionic current runs, as well as inside an ion-exchange membrane (IEM). In addition, there is electrostatic heating and cooling in the nanoscopic regions on each outer surface of the IEM, i.e., in the Donnan layers, or electrical double layers (EDLs), that form there, in which the ion concentrations sharply change from their values in the open electrolyte fluid just outside the membrane, to their values just within the IEM. At the side of the membrane where the current $I_{mem}$ (directed into the IEM) and the Donnan potential $\Delta\phi_D$ (defined as the potential in the membrane minus that just outside) have opposite signs, there will be electrostatic heating in the EDL. And by the same argument, on the other side of the membrane, thus where $I_{mem}$ and $\Delta\phi_D$ are of equal sign, there is electrostatic *cooling*. Heat will flow to this location to reduce the temperature difference with the surrounding. Thus, in any ion-exchange membrane process, with current running through the membrane, besides the Joule heating within the membrane and in the outer electrolyte phases (open channels, unstirred layers), there is additional heating directly on one interface of the membrane (in the Donnan layer there), while there is cooling on the other interface of the membrane. This is similar to the Peltier effect where upon directing an electrical current through two electrical potential discontinuities (one up, one down), at one point heat is produced, and at the other point heat is removed [ref. 9, p.191].

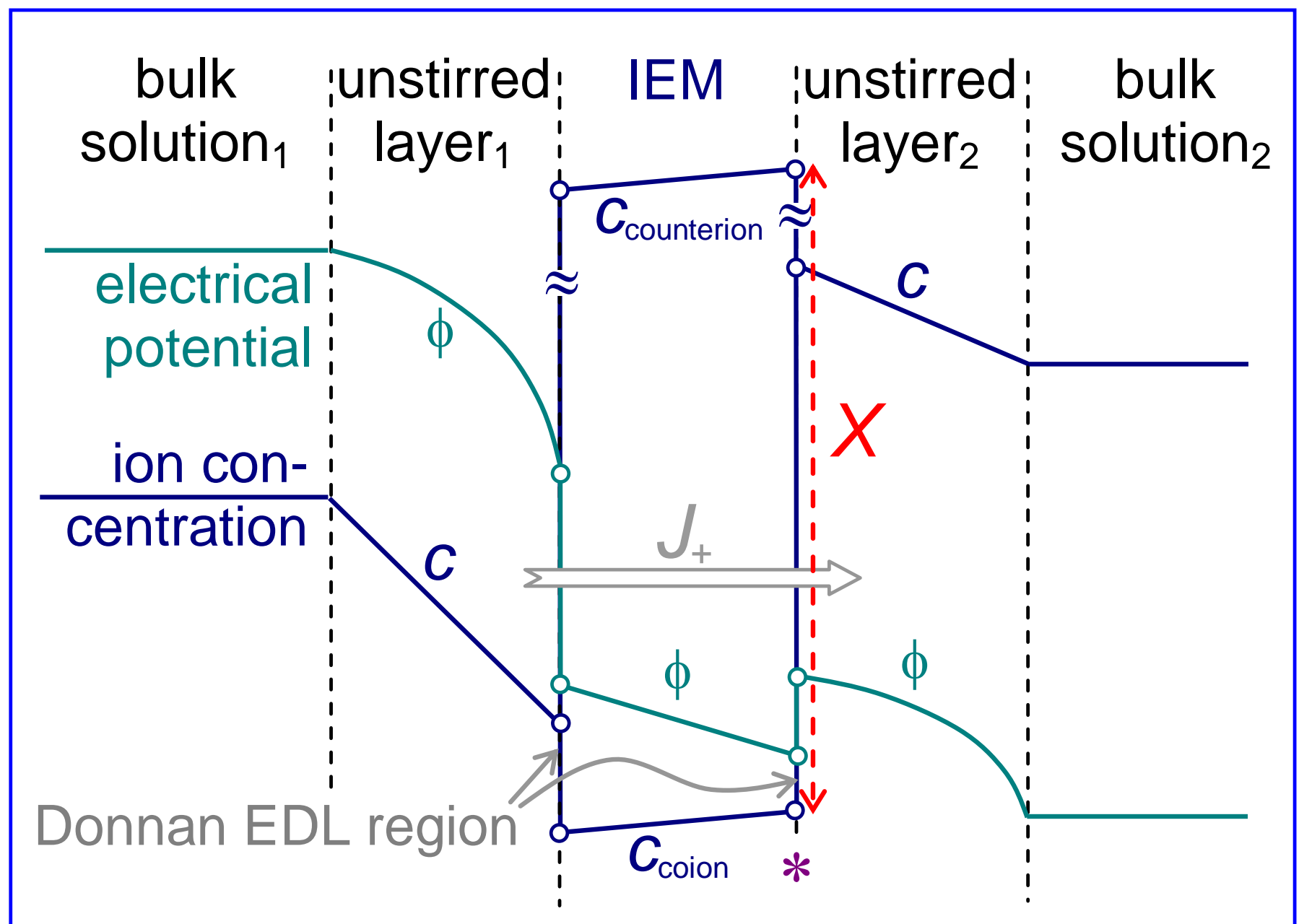

Fig. 1. Schematic of an ion-exchange membrane (IEM) in between two electrolyte phases. Assuming the IEM is a cation-exchange membrane (CEM), the counterions are cations, the coions are anions. The fixed membrane charge density is denoted by $X$. If there is a current running left to right in a CEM, it will be mainly carried by the cation flux, $J_+$, which is directed to the right in the figure. In such a process, there will be Joule heating in all bulk phases, where the electrical potential $\phi$ decreases towards the right. However, at positions where the potential *in*creases in that direction, namely in the EDL on the right-hand of the membrane (shown with *), there is electrostatic cooling.

To understand electrostatic cooling, it must be realized that current not always flows exactly down the potential gradient. Instead, both the current **I** and the field strength **E** are vectors and may point in different directions, and it is the inner product of these vectors, **I·E,** which is the electrical power density that enters the heat equation (thermal energy balance) [ref. 9, p. 197; ref. 10, p. 93; ref. 11, Eq. 14; ref. 12, p. 323; ref. 13, p. 21; ref. 14]. The larger the angle between **I** and **E**, the smaller is the heating. When the two vectors are at right angles, there is no heating. Continuing the same argument, when the angle between the two vectors is >90°, there must be cooling. When the two vectors are in the exact opposite direction, the heat production is equal to minus the product of the norm of **I** and the norm of **E**, which is a negative scalar quantity, see Fig. 2.

For the dynamic formation of an EDL at a metal electrode, it was already pointed out by Kontturi *et al.* [13, p. 32] that **I**·**E** may be less than zero, leading to a decrease in internal energy. For the same problem of EDL charging, d'Entremont and Pilon [14] develop the term **I**·**E** by implementing the Nernst-Planck equation (and modifications thereof) to arrive at the strictly positive term $q_{E,j}=\mathbf{I}^2/\sigma$ which is the Joule heating (with $\sigma$ the conductivity), as one of several contributions in the thermal energy balance, with other terms that include $q_{E,d}=D\sigma^{-1}\mathbf{I}\cdot\nabla\rho_e$, where $\rho_e$ is the local charge density due to the mobile ions. This form of the extra term $q_{E,d}$ is obtained from the Nernst-Planck equation for a symmetric ion mixture with ions of equal valency and diffusion coefficients and is modified to $q_{E,d}=\sigma^{-1}\mathbf{I}\cdot\sum_i z_i D_i \nabla c_i$ in case of mixtures of ions with unequal valencies and diffusion coefficients. In a very different field, Robson *et al.* [15] describe how a current flow opposing the electric field in a plasma also leads to cooling.

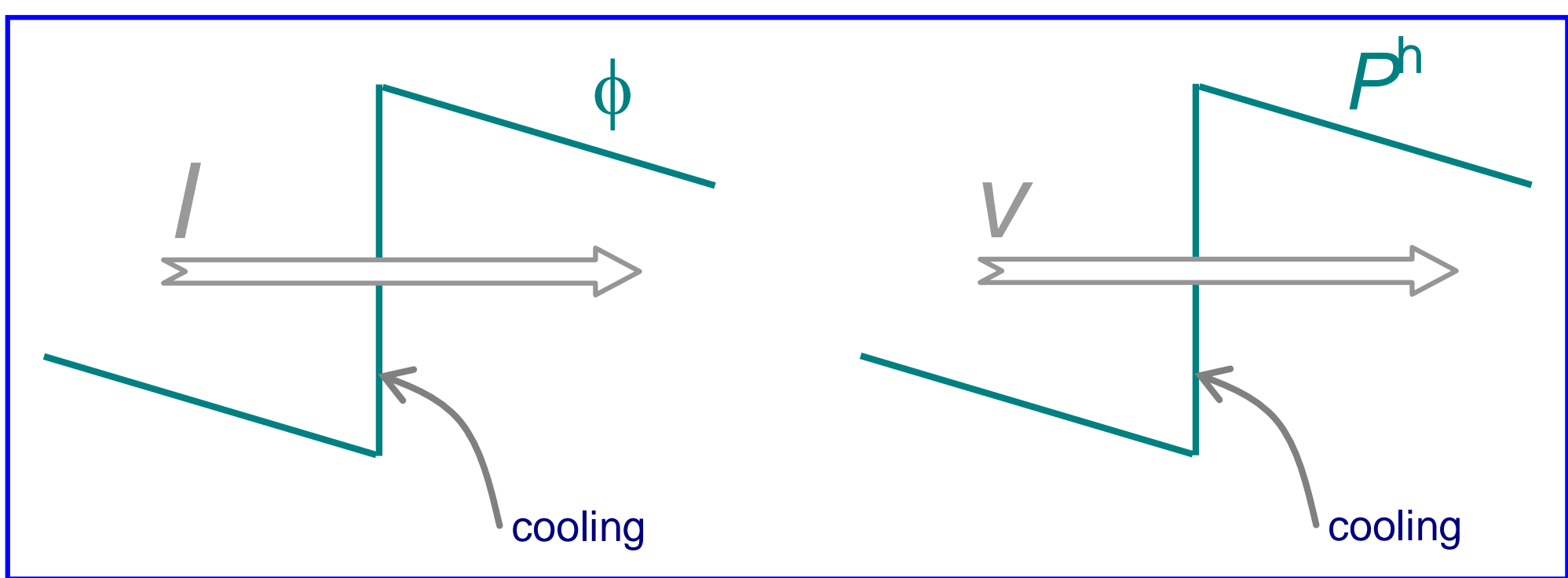

Fig. 2. Local cooling when current flows to a higher electrical potential, or when fluid flows to a higher hydrostatic pressure.

In most IEM processes, ionic current is driven by an externally applied voltage across a system of alternating cation-exchange membranes, CEMs, and anion-exchange membranes, AEMs, as in electrodialysis (ED), see Fig. 1. In such a process, the cooling in the EDL on the concentrate side of the membranes will be smaller in magnitude than the heating on the dilute side (both for the CEMs and for the AEMs). On top of that, there is Joule heating in the bulk of the membranes and across the spacer channels, and, thus, there is a net production of heat in the entire system. However, in the reverse process, called reverse electrodialysis (RED), where a natural difference in salt concentration, for instance between river water and seawater, drives the spontaneous generation of current, the situation is reversed, and there is cooling in the EDL on the dilute side of the membrane (both for CEMs and AEMs), which is larger in magnitude than the heating in the EDL on the concentrate side of the membrane. In addition, as long as electrical power is generated by a RED system, this cooling is in magnitude larger than the sum of all of the Joule heating occurring in the bulk of membranes, the spacer channels, unstirred layers, and the electrostatic heating in the EDL on the concentrate side of the membrane combined. This must be the case because the production of electrical power in RED implies a total cooling of the water streams of the same magnitude, assuming ideal salt solutions and neglecting the heating due to the pumping of the solution through the channels (viscous dissipation). It should be possible to measure the temperature decrease of the salt solutions between inlet and outlet in a RED system, or measure the difference in temperature between each side of one of the ion-exchange membranes.

Let us reiterate the above arguments by considering a RED "stack" without fluid flow: the aqueous solutions are stagnant and thus we can neglect viscous dissipation by fluid flow and associated heat effects. Thus we have stagnant fluid layers between pairs of membranes, where alternatingly one fluid is dilute and the other is concentrated in salt. In this planar one-dimensional geometry the array of AEMs and CEMs forms the membrane stack. Let us assume the presence of capacitive (porous carbon) electrodes on the two ends [16]. The spontaneous ionic movement that will develop, with in each membrane counterions running from the concentrated to the dilute compartments, will set up an ionic current, which translates into an electrical current in the external circuit, which we can direct through a resistance which heats up. As the thermal energy balance for the entire system must be zero when the salt solutions are thermodynamically ideal, somewhere the salt solutions or membranes must cool off (to compensate for the heating of the resistance). The total cooling in the entire aqueous phase (channels, membranes) is equal to the ionic current times the cell voltage difference between the two electrodes (integrated over time). Then the question is, where exactly in this system is heat produced and where is heat absorbed? To answer this question, it is quite logical then to assume that the local cooling/heating rate is described by the ionic current times the local potential gradient, not only in the bulk of the solutions and the membranes (where it will always lead to heating), but also in the EDLs formed on the membrane/water interface (where we argue that on the dilute side the interface will cool). This assumption for the local heat generation exactly matches the macroscopic observation that when electrical power is generated from salt concentration differences, there is not only local heating within the membranes and in the aqueous solutions, but there must also be local cooling somewhere, as overall there must be a cooling of the system (channels, membranes) equal in magnitude to the generated electrical power which is the product of cell voltage and current.

The same phenomenon of electrostatic cooling will occur in any EDL through which currents run down a potential gradient (leading to heating), or up (leading to cooling), either, as discussed above, for a process such as (R)ED for steady-state EDLs formed on each side of a membrane, but just as well in case of dynamically forming (transient) EDLs, for instance in porous electrodes for EDL-capacitors, capacitive deionization, and capmix [14,17-19].

When instead of a current, or in addition to a current, also solvent is transported through the membrane, a similar effect can be expected to occur. It is well-known that when solvent (or fluid in general) flows down a hydrostatic pressure gradient, $\nabla P^h$, that the product of volumetric flow rate (fluid velocity), $\mathbf{v}$, and $-\nabla P^h$ leads to heat production, i.e., viscous dissipation. We argue now that whenever there is fluid flow towards a higher hydrostatic pressure, there must be pressure-related cooling. Thus, in a membrane process with fluid flow through the membrane, there will be locations with pressure-related heating, and other locations with cooling. Using a charged membrane, the osmotic pressure will be larger in the membrane than just outside the membrane, and thus the hydrostatic pressure, just like the osmotic pressure, makes a distinct upward jump from outside to inside [3,7,20,21]. Thus, pressure-related heating in an EDL takes place on that side of the charged membrane where the water exits the membrane, while there is cooling in the EDL where the water enters the membrane. This situation occurs for any membrane process with fluid flow, such as reverse osmosis, forward osmosis and pressure-retarded osmosis, where the use of charged membranes implies that the osmotic pressure inside the membrane is larger than just outside. However, for pure hyperfiltration membranes where (often, neutral) constituents in the water are rejected from the membrane by a sieving mechanism [21,22], the situation will be the opposite: the osmotic and hydrostatic pressure is lower in the membrane than just outside, and pressure-driven cooling occurs on the side where the fluid exits the membrane. Thus, by measurement of the heat production at the interface of a membrane subject to fluid flow, the osmotic pressure difference across the water/membrane interface can be calculated.

Theoretically, pressure-related cooling can be included in the thermal energy balance of ref. 11 (Eq. 14) to arrive at

$$\rho C_{\mathrm{p}}\left(\frac{\partial T}{\partial t}+\mathbf{v}\cdot\nabla T\right)=\lambda\nabla^{2}T+\mathbf{I}\cdot\mathbf{E}-\mathbf{v}\cdot\nabla P^{\mathrm{h}}\,. \tag{1}$$

Eq. (1) can also be derived from Eq. 187, p. 129, in ref. 23, which can be written as

$$\frac{\partial u}{\partial t}=-\nabla\left(\mathbf{v}\cdot P^{\mathrm{h}}+\mathbf{J}_{\mathrm{q}}+\mathbf{I}\cdot\phi\right)+\rho_{\mathrm{e}}\frac{\partial\phi}{\partial t} \tag{2}$$

where $\rho_e$ is the local charge density (C/m$^3$). Eq. (2) can be rewritten to Eq. (1) considering the specific energy $u=\rho_{\mathrm{m}}C_{\mathrm{p}}\left(T-T_{\mathrm{ref}}\right)+\rho_{\mathrm{e}}\phi$, the heat flux $\mathbf{J}_{\mathrm{q}}=\rho_{\mathrm{m}}C_{\mathrm{p}}\left(T-T_{\mathrm{ref}}\right)\mathbf{v}-\lambda\nabla T$, and continuity of mass and charge, $\nabla\cdot\mathbf{v}=0$ and $\frac{\partial\rho_{\mathrm{e}}}{\partial t}=-\nabla\cdot\mathbf{I}$. Note that **I** is the ionic conduction current (Eq. 185, p. 129, in ref. 23).

Besides ion-exchange membranes, we also expect these negative contributions to the thermal energy balance to occur in charged nanochannels and microchannels [24-26] when driven by salt concentration gradients across the channel, currents and fluid flows (or both) are generated against their associated forces, potential and pressure gradients.

It must be noted that the temperature decrease is expected to be small. For instance, in water, considering the EDL at the membrane/solution-interface, based on a thermal conductivity of $\lambda$=0.6 W/m/K, and assuming a typical boundary layer thickness for heat transfer of $L$=100 $\mu$m, and with an assumed Donnan potential of $\Delta\phi_D$=0.5 V and a current density of $I$=100 A/m$^2$, we arrive at $\Delta T=I\cdot\Delta\phi_D/(\lambda\cdot L)$~8 mK. For a membrane process with fluid flow, such as reverse osmosis, assuming over the EDL an osmotic pressure difference (thus, hydrostatic pressure difference) of $\Delta P$=50 bar (e.g., because just outside the membrane the concentration of ions is 1 M ions, and inside it is 3 M), and assuming heat to be transported away convectively, then with a volumetric heat capacity of water of $\rho_m C_p$=4.2 MJ/m$^3$/K, the temperature change is $\Delta T=\Delta P/\rho C_p$~1.2 K, a number about 100x larger.

**Acknowledgments**

The authors thank Joost Veerman (REDstack B.V., Sneek, Netherlands) for many useful and relevant suggestions which improved the manuscript.